\documentstyle[12pt]{article}

\newcommand{\AmS}{{\protect\the\textfont2
\renewcommand{\thesection}{\Roman{section}}
  A\kern-.1667em\lower.5ex\hbox{M}\kern-.125emS}}
 
\begin{document}
\rightline {CERN-TH/96-194}
\vskip 2. truecm
\centerline{\bf Complex Probability Distributions: A Solution for the }
\centerline{\bf Long-Standing Problem of QCD at Finite Density }
\vskip 2 truecm
\centerline { Vicente ~Azcoiti}
\vskip 1 truecm
\centerline {\it CERN Theory Division,}
\centerline {\it CH-1211, Geneve 23, Switzerland}
\vskip 3 truecm
\centerline {ABSTRACT}
\vskip 0.5 truecm
We show how the prescription of taking the absolute value of the 
fermion determinant in the integration measure of QCD at finite density, 
forgetting its phase, 
reproduces the correct thermodynamical limit. This prescription, which 
applies also to other gauge theories with non-positive-definite 
integration measure, also has the advantage of killing finite size 
effects due to extremely small mean values of the cosine of the phase 
of the fermion determinant. We also give an explanation for the pathological 
behaviour of quenched QCD at finite density.

\vfill
\begin{flushleft}
CERN-TH/96-194\\
July 1996
\end{flushleft}
\eject
\par
\leftline{\bf 1. Introduction}
\par
Non-perturbative investigations of QCD at finite temperature and density 
have received much attention in the last years. The aim of these 
investigations is to find the matter conditions in the early Universe 
and to get a clear insight into experimental signatures in the heavy-ion 
collision experiments. Even if considerable progress has been achieved 
in the investigations of QCD at finite temperature and zero chemical 
potential using the lattice approach, the present situation of the field at 
finite density is not so satisfactory. As is well known, the complex nature 
of the determinant of the Dirac operator at finite chemical potential, 
which makes it impossible to use standard simulation algorithms based on 
positive-definite probability distribution functions, has much delayed 
investigations on the full theory with dynamical fermions. On the other 
hand the quenched approximation, which has been extensively and successfully 
used in simulations of QCD at zero chemical potential, seems to have 
some pathological behaviour when applied to QCD at finite density 
\cite{BAR,KOG}.

I want to show in this article how one of the main technical difficulties 
in simulating QCD at finite density, the complex fermion determinant, can 
be easily surmounted. In fact I will demonstrate that it is enough,  
in order to get 
the correct thermodynamical limit to take the absolute 
value of the fermion determinant in the integration measure, forgetting 
completely the contribution of its phase. 
I will also show how not only do we get the correct 
thermodynamical limit in this way but also that it is an efficient way to 
kill finite size effects present in the exact simulations and related to 
very small expectation values of the cosine of the fermion 
determinant phase \cite{TOUS}.

This result, which applies also to more general 
cases such as non-positive-definite fermion determinants in gauge theories 
with Wilson fermions \cite{SCHWIL}, 
agrees with a recent finding of Stephanov 
\cite{MISHA} in the random matrix model approximation. In fact QCD 
at finite chemical potential and with $n$ flavours can be seen as a 
theory with $\frac{n} {2}$ quarks with original action and 
$\frac{n} {2}$ with conjugate action. This notwithstanding, the zero 
flavour limit of the model corresponds precisely to standard quenched 
QCD at finite density. The anomalous behaviour of the quenched model 
observed in the numerical simulations remains therefore unclear yet. 
We will give in this paper an 
explanation for the pathological behaviour of the quenched 
model in the forbidden region, a region of the chemical potential 
$\mu$ characterized by wild fluctuations \cite{KOGKOG}.

\vfill
\eject
\leftline{\bf 2. Complex Distributions}
\par
Let me fix here the standard notation in gauge theories with dynamical 
fermions, even if the results I am going to discuss apply to every 
equilibrium statistical mechanics system with a complex "Boltzmann weight".
The starting point is the following partition function 

$$
Z= \int [dU] e^{-\beta S_{G}(U)} \det \Delta(U,m,\mu)
\eqno(1)
$$
\noindent
where $U$ are the gauge variables (elements of $SU(3)$ in the QCD case), 
$\beta$ the inverse gauge coupling, $S_G(U)$ the pure gauge action and 
$\Delta(U,m,\mu)$ the lattice Dirac operator at fermion mass $m$ and 
chemical potential $\mu$.

There are several physically interesting cases in which the determinant of 
the Dirac operator $\Delta(U)$ for a generic 
gauge configuration $U$ is either 
not positive-definite (gauge theories with Wilson fermions) or even 
a complex number (QCD at finite density). In all these cases, standard 
simulation algorithms based on the interpretation of the fermion 
determinant as a probability distribution function $(p.d.f.)$ 
to be multiplied by the 
pure gauge probability distribution $e^{-\beta S_{G}(U)}$ fail. The 
standard way to overcome this problem is to take the absolute value of 
the fermion determinant in the $p.d.f.$ of the path integral. The integration 
measure becomes

$$
[dU] e^{-\beta S_{G}(U)} \left |\det \Delta(U,m,\mu)\right |.
\eqno(2)
$$

The vacuum expectation value of any operator $O(U)$ with the previous 
prescription now becomes 

$$
{\langle\,O(U)\rangle} = 
\frac{{\langle\,O(U)e^{i\phi_{\Delta}}\rangle}_{||}} 
{{\langle\,e^{i\phi_{\Delta}}\rangle}_{||}}
\eqno(3)
$$

\noindent
where $\phi_{\Delta}(U,m,\mu)$ is the phase of the determinant of the 
Dirac operator and 
$\langle\ \rangle_{||}$ in (3) represents the mean value computed with 
the $p.d.f.$ (2).

In QCD at finite chemical potential and due to the symmetries of the 
action, the numerator and denominator of (3) are real numbers. However 
simulations of this model show that the real part of the denominator of (3) 
in the physically interesting region of the chemical potential 
becomes extremely small and impossible to measure numerically \cite{TOUS}. 
As I will show, this kind of measurements are not necessary 
since in the thermodynamic limit we get the following factorization:

$$
{\langle\,O(U)e^{i\phi_{\Delta}}\rangle}_{||} = 
{\langle\,O(U)\rangle}_{||}{\langle\,e^{i\phi_{\Delta}}\rangle}_{||}
\eqno(4)
$$

\noindent
for any intensive operator $O(U)$. Equation (4) implies that by taking 
the absolute value prescription in the 
integration measure instead of the fermion determinant we get the correct 
thermodynamical limit, i.e. 

$$
\lim_{V\rightarrow\infty} {\langle\,O(U)\rangle} = 
{\langle\,O(U)\rangle}_{||}.
\eqno(5)
$$

To show the correctness of equation (5) let me consider the partition 
function (1) and write it as 

$$
Z = {\langle\,e^{i\phi_{\Delta}}\rangle}_{||}
\int [dU] e^{-\beta S_{G}(U)} \left |\det \Delta(U,m,\mu)\right |.
\eqno(6)
$$

\noindent
The vacuum expectation value of any thermodynamical quantity, the chiral 
condensate for instance, can be written as

$$
{\langle\,\bar\psi\psi\rangle} = 
\lim_{V\rightarrow\infty} \frac{1} {V} Z^{-1} 
\frac{\partial Z} {\partial m} $$
$$
= 
\lim_{V\rightarrow\infty} \frac{1} {V} \left(Z_{||}^{-1} \frac{\partial Z_{||}} 
{\partial m} + {\langle\,e^{i\phi_{\Delta}}\rangle}_{||}^{-1} 
\frac{\partial {\langle\,e^{i\phi_{\Delta}}\rangle}_{||}} {\partial m}\right)
\eqno(7)
$$
\noindent
with 

$$
Z_{||} = \int [dU] e^{-\beta S_{G}(U)} \left |\det \Delta(U,m,\mu)\right |.
\eqno(8)
$$

Since ${\langle\,e^{i\phi_{\Delta}}\rangle}_{||}$ is a bounded function 
of the system's parameters for every lattice volume, it takes a finite 
value in the infinite volume limit. Therefore the second contribution 
to the expectation value of equation (7) will vanish in the thermodynamical 
limit except, at most, in some isolated points. It gives only non-vanishing 
values at finite volume. These are pure finite size effects but they  
can significantly distort the results on finite lattices in regions of the 
parameters where this term could be large \cite{TOUS}. 

These results apply for any thermodynamical quantity. The general rule is 
therefore to take $Z_{||}(\beta,m,\mu)$ as the generating partition 
function, the logarithmic derivatives of which will give us the right 
vacuum expectation values. Notice also that the practical rule of taking 
the absolute value of the fermion determinant works also for any 
intensive operator, like correlation functions, which can be obtained as 
a derivative of the partition function with external sources.

\vfill
\eject
\leftline{\bf 3. The Quenched QCD Puzzle}
\par
The results of the preceding section tell us that the fermionic contribution 
to the integration measure of QCD at finite chemical potential can be 
written as $(\det\Delta \det\Delta^{+})^{1/2}$, i.e. QCD with $n$ dynamical 
flavours is a theory with $\frac{n} {2}$ quarks with original action and 
$\frac{n} {2}$ quarks with conjugate action \cite{MISHA}. The zero flavour 
limit of this model is however standard quenched QCD. I will show here, 
with the help of the fermion effective action formalism \cite{TRIO}, that 
quenched QCD at small but finite chemical potential actually breaks 
dynamically chiral symmetry. Furthermore the chiral transition at finite 
$\mu$ is second order in the quenched model, a fact that is most likely to be  
a pathology of the quenched approximation and which will be removed with 
the inclusion of dynamical fermions. The very large fluctuations observed 
in quenched simulations \cite{KOGKOG} 
should be a manifestation of the second-order 
character of the phase transition. If the inclusion of dynamical fermions 
makes the phase transition of first order, as expected, fluctuations will 
decrease due to a strong selection in the relevant configuration sample 
caused by the inclusion of the fermion determinant in the integration 
measure.

The effective fermion action formalism is based on the definition of 
an effective fermion action, 
which depends on the gauge energy density $E$, 
bare fermion mass $m$ and chemical potential $\mu$. This can be done by 
including in expression (8) a $\delta$ function 
$\delta(\frac{1} {6V} S_G -E)$ and a integral over the gauge 
energy density $E$ \cite{TRIO}. 
This allows us to write the partition function (8) as 
a one-dimensional integral: 

$$
Z= \int dE N(E) e^{-6\beta VE} 
{\langle\,\left |\det \Delta(U,m,\mu)\right |\rangle}_E.
\eqno(9)
$$

\noindent
$N(E)$ is the density of states of fixed energy $E$ and 
${\langle\,\rangle}_E$ the mean value computed over gauge configurations 
of fixed energy density $E$. The normalized fermion effective action 
$S_{eff}^{F}(E,m,\mu)$ is then defined as 

$$
{S_{eff}^{F}(E,m,\mu)} = 
- \frac{1} {V} \log {\langle\,\left |\det \Delta(U,m,\mu)\right |\rangle}_E.
\eqno(10)
$$

The thermodynamics of this system can be solved in the infinite volume 
limit by the saddle point technique. The chiral condensate 
${\langle\,\bar\psi\psi\rangle}$ and number density 
${\langle\,J_{0}\rangle}$ will be respectively given by

$$
{\langle\,\bar\psi\psi\rangle}= 
-\frac{\partial S_{eff}^{F}} {\partial m}\;\;\;\;\;\;\;
{\langle\,J_{0}\rangle}= 
-\frac{\partial S_{eff}^{F}} {\partial \mu}, 
\eqno(11)
$$

\noindent
both expressions evaluated at the energy $E(\beta,m,\mu)$ which 
satisfies the saddle point equation. In the quenched approximation the 
fermion effective action does not appear in the integration measure. The 
saddle point solution for the plaquette energy $E$ depends only on the 
inverse gauge coupling $\beta$ in this approximation; therefore, it does not 
change by changing the fermion mass $m$ or the chemical potential $\mu$. 
The fact that also in the quenched approximation the chiral condensate and 
number density are finite numbers for any value of the gauge coupling 
$\beta$ tells us that the fermion effective action must be a continuous 
function of $m$ and $\mu$ for every value of $E$ and with finite first 
derivatives. Simple mathematics shows that also 
$\frac{\partial S_{eff}^{F}} {\partial m}$ and 
$\frac{\partial S_{eff}^{F}} {\partial \mu}$ will then 
be continuous functions 
of $\mu$ and $m$, respectively. Taking now into account that 
$\frac{\partial S_{eff}^{F}} {\partial m}$ is the chiral condensate and that 
by changing $\mu$ the energy $E$ does not change in the quenched 
approximation, we get as a result that the chiral condensate is a continuous 
function of $\mu$ for every value of $m$ in this approximation. Since 
chiral symmetry is dynamically broken at $\mu=0$ it must be broken also 
at small $\mu$. Even more, the chiral transition at finite $\mu$ must 
be continuous. The only way to get a discontinuous transition is to have 
a two-minimum structure in the full effective action (9). The compact $U(1)$ 
model has a first order chiral transition in the quenched approximation, 
since the pure gauge action has a two-minimum structure. 
Since the pure gauge $SU(3)$ model 
has no first order transitions at zero temperature, only 
the inclusion of dynamical fermions can produce such a structure 
in the full effective action.
In such a case and for some selected values of the model parameters, the 
plaquette energy, which verifies the saddle point equation, will jump 
between two different values as well as the other thermodynamical 
quantities.

\vskip 1truecm
\leftline{\bf 4. Summary}
\par
I have shown here how the difficulty in applying standard simulation 
algorithms to QCD at finite density, due to the complex nature of the 
fermion determinant, can be easily surmounted by taking the absolute 
value in the integration measure. This prescription not only gives 
the properly behaved thermodynamical limit but also has the advantage 
to kill unwanted finite size effects. I have also shown that the 
solution to the quenched QCD puzzle is on the line 
pointed out in \cite{KOGKOG}, 
i.e. chiral symmetry is spontaneously broken at small $\mu$. The only 
pathology of the quenched approximation, which is most likely to change 
with the inclusion of dynamical fermions, is the continuous character 
of the chiral transition; this allows us to understand the large 
fluctuations observed in quenched simulations.

\vfill
\eject


\vskip 1 truecm
\begin{thebibliography}{9}

\bibitem{BAR}
I. Barbour, N. Behilil, E. Dagotto, F. Karsch, A. Moreo, M. Stone, 
H.W.~Wyld, 
Nucl. Phys. {\bf B275 [FS17] } \rm (1986) 296. 

\bibitem{KOG}
J.B. Kogut, M.P. Lombardo, D.K. Sinclair, 
Phys. Rev. {\bf D51} \rm (1995) 1282.

\bibitem{TOUS}
D. Toussaint, 
Nucl. Phys. {\bf B17 } Proc. Suppl. \rm (1990) 248. 

\bibitem{SCHWIL}
V. Azcoiti, G. Di Carlo, A. Galante, A.F. Grillo, V. Laliena, 
Phys. Rev. {\bf D53} \rm (1996) 5069; P. de Forcrand, Proceedings of 
the Lattice 96 Symposium, 
Nucl. Phys. {\bf B } Proc. Suppl. \rm to appear. 

\bibitem{MISHA}
M.A. Stephanov, 
hep-lat {\bf $9604003$} \rm (1996), to be published in Phys. Rev. Lett.

\bibitem{KOGKOG}
M.P. Lombardo, J.B. Kogut, D.K. Sinclair, 
hep-lat {\bf $9511026$} \rm November (1995).

\bibitem{TRIO}
V. Azcoiti, G. Di Carlo, A.F. Grillo, 
Phys. Rev. Lett. {\bf 65} \rm (1990) 2239. 


\end{thebibliography}
\end{document}